\documentclass[a4paper,11pt]{article}
\usepackage{pos}

\usepackage{subcaption}
\usepackage{units}
\usepackage{gensymb}
\usepackage{mathrsfs}
\usepackage{mathtools}
\usepackage{bm}

\def\Journal#1#2#3#4{{#1} {\bf #2}, #3 (#4)}

\def\ADASSA{\em Proceedings of XXX Astronomical Data Analysis Software and Systems (ADASS) conference}
\def\ADASSB{\em Proceedings of XXXI Astronomical Data Analysis Software and Systems (ADASS) conference}
\def\APJ{\em Astrophys. J.}
\def\ASPP{\em Astroparticle Physics}
\def\CVPR{\em Proceedings of the IEEE conference on computer vision and pattern recognition}

\def\ICASSP{\em Proceedings of IEEE International Conference on Acoustics, Speech and Signal Processing (ICASSP)}
\def\ICRCA{\em Proceedings of $36^{\text{th}} $ International Cosmic Ray Conference (ICRC)}
\def\ICRCB{\em Proceedings of $37^{\text{th}} $ International Cosmic Ray Conference (ICRC)}
\def\ICRCC{\em Proceedings of $33^{\text{th}} $ International Cosmic Ray Conference (ICRC)}
\def\ICRCD{\em Proceedings of $35^{\text{th}} $ International Cosmic Ray Conference (ICRC)}

\def\ZENODO{\em Zenodo}

\title{The performance of the MAGIC telescopes using deep convolutional neural networks with CTLearn}
\ShortTitle{MAGIC deep learning analysis with CTLearn}

\author*[a]{T. Miener}
\author[a]{D. Nieto}
\author[b]{R. L\'{o}pez-Coto}
\author[a]{J. L. Contreras}
\author[c]{J. G. Green}
\author[c]{D. Green}
\author[d]{E. Mariotti}
\onbehalf{on behalf of the MAGIC Collaboration}

\affiliation[a]{Instituto de F\'{i}sica de Part\'{i}culas y del Cosmos and Departamento de EMFTEL, Universidad Complutense de Madrid, Spain}
\affiliation[b]{Instituto de Astrofísica de Andalucía - CSIC, Granada, Spain}
\affiliation[c]{Max-Planck-Institut für Physik, München, Germany}
\affiliation[d]{Dipartimento di Fisica e Astronomia dell'Università and Sezione INFN, Padova, Italy}

\emailAdd{tmiener@ucm.es}

\abstract{The Major Atmospheric Gamma Imaging Cherenkov (MAGIC) telescope system is located on the Canary Island of La Palma and inspects the very high-energy (VHE, few tens of GeV and above) gamma-ray sky. MAGIC consists of two imaging atmospheric Cherenkov telescopes (IACTs), which capture images of the air showers originating from the absorption of gamma rays and cosmic rays by the atmosphere, through the detection of Cherenkov photons emitted in the shower. The sensitivity of IACTs to gamma-ray sources is mainly determined by the ability to reconstruct the properties (type, energy, and arrival direction) of the primary particle generating the air shower. The state-of-the-art IACT pipeline for shower reconstruction is based on the parameterization of the shower images by extracting geometric and stereoscopic features and machine learning algorithms like random forest or boosted decision trees. In this contribution, we explore deep convolutional neural networks applied directly to the pixelized images of the camera as a promising method for IACT full-event reconstruction and present the performance of the method on observational data using \texttt{CTLearn}, a package for IACT event reconstruction that exploits deep learning.}

\FullConference{%
  7th Heidelberg International Symposium on High-Energy Gamma-Ray Astronomy (Gamma2022)\\
  4-8 July 2022\\
  Barcelona, Spain\\}

\begin{document}
\maketitle

\section{Introduction}
In this contribution, we show how deep convolutional neural networks (CNNs) can be utilized to detect astrophysical gamma-ray sources like the Crab Nebula using \texttt{CTLearn}\footnote{\href{https://github.com/ctlearn-project/ctlearn}{https://github.com/ctlearn-project/ctlearn}}~\cite{Brill:CTLearn,2017arXiv170905889N,2019ICRC...36..752N,2022ASPC..532..191N}, a deep learning (DL) framework for IACT event reconstruction, and \texttt{DL1-Data-Handler}\footnote{\href{https://github.com/cta-observatory/dl1-data-handler}{https://github.com/cta-observatory/dl1-data-handler}} (\texttt{DL1DH})~\cite{Kim:DL1DH}, a package designed for the data management of machine learning image analysis techniques for IACT data. The results are compared to the standard analysis (random forest (RF) for the background rejection, Look-Up tables (LUTs) for the energy estimation and RF for bidimensional direction reconstruction) obtained with MAGIC Analysis and Reconstruction Software \texttt{MARS}~\cite{Zanin:MARS,Aleksic:MAGIC}. Previous DL analyses of MAGIC data~\cite{2021arXiv211201828M} were carried out with \texttt{CTLearn v0.5} based on \texttt{TensorFlow}\footnote{\href{https://www.tensorflow.org/}{https://www.tensorflow.org/}}~\texttt{v1}, while this work used \texttt{CTLearn v0.6}, which adopted the Keras\footnote{\href{https://keras.io/}{https://keras.io/}} API~\cite{Chollet:Keras} from \texttt{TensorFlow v2}~\cite{tensorflow_developers_2022_5949125}.

\begin{figure}[ht]
    \centering
    \includegraphics[width=1.0\textwidth]{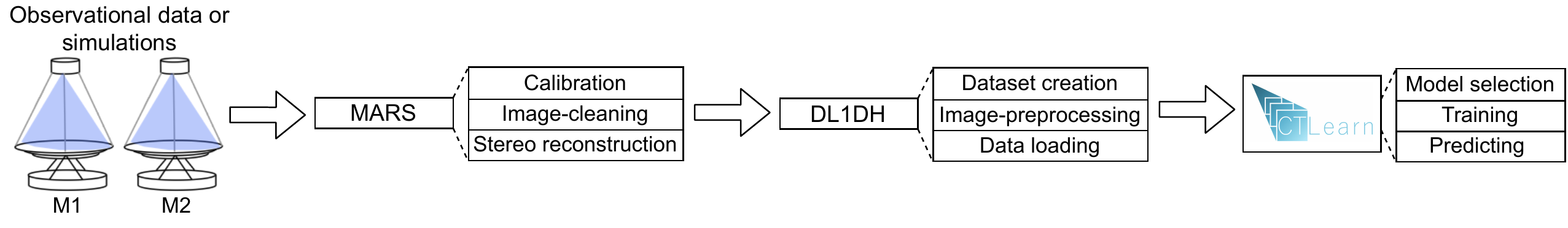}
    \caption{Workflow of the MAGIC DL analysis with \texttt{CTLearn}~\cite{2021arXiv211201828M}.}
    \label{fig:MAGICDLWorkflow}
\end{figure}

The workflow of the MAGIC DL analysis with \texttt{CTLearn} is illustrated in Fig.~\ref{fig:MAGICDLWorkflow}. First, the images are calibrated and cleaned by \texttt{MARS} to suppress the major fraction of the Night Sky Background (NSB). Crucial information are translated into \texttt{uproot}\footnote{\href{https://github.com/scikit-hep/uproot4}{https://github.com/scikit-hep/uproot4}}-readable branches~\cite{Pivarski:uproot} using a complementary macro. Then, the \texttt{DL1DH} assembles several data levels from \texttt{MARS} and unifies them in a common data format in \texttt{HDF5} designed for DL purposes. The image preprocessing and data reading is managed by the \texttt{DL1DH}. Bilinear interpolation is used to map the hexagonal pixel layout of the MAGIC cameras to a Cartesian lattice to directly apply CNNs~\cite{2019ICRC...36..753N}. Finally, \texttt{CTLearn} performs training and prediction with CNN-based models, allowing for full-event reconstruction.

\section{DL analysis with the MAGIC telescopes}

\subsection{Model selection}
\begin{figure}[ht]
    \centering
    \includegraphics[width=0.45\textwidth]{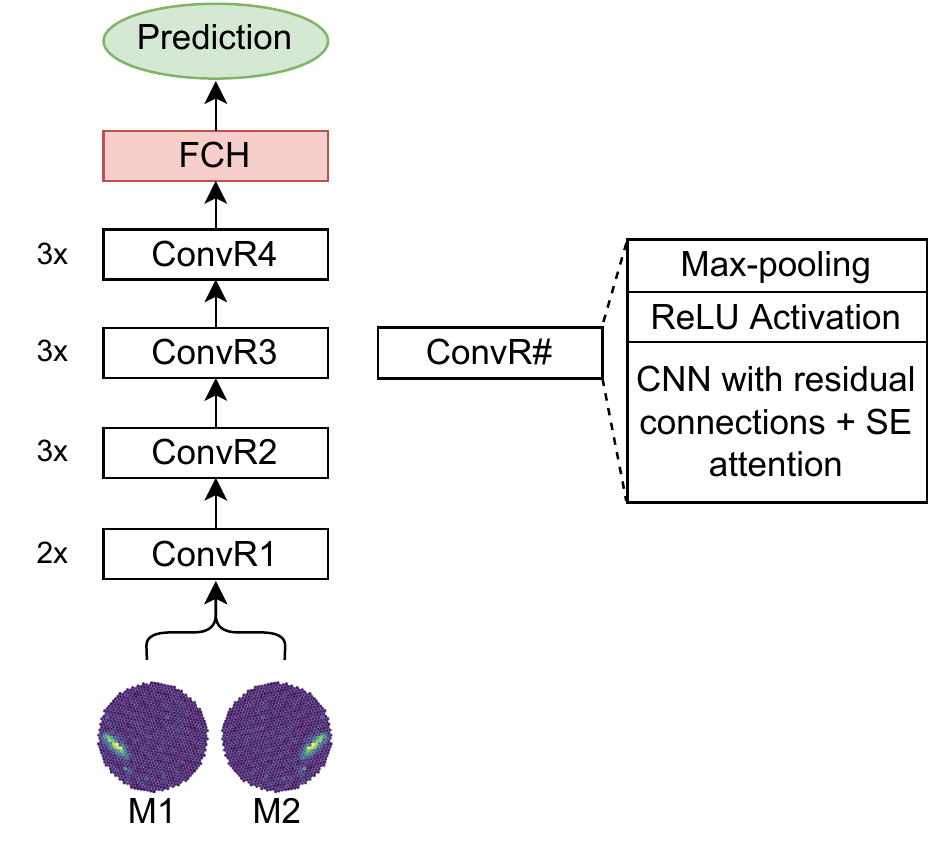}
    \caption{\texttt{CTLearn}’s TRN model with channel-wise concatenation of the two stereoscopic images recorded by the MAGIC telescopes (M1 and M2).}
    \label{fig:MAGICDLTRNmodel}
\end{figure}

For this work, we selected \texttt{CTLearn}’s Thin-ResNet (TRN)~\cite{2019arXiv190210107X} model~\cite{Grespan:LST,Miener:CTA}, which is a shallow residual neural network~\cite{He:ResNet} with 33 layers\footnote{The first initialization layer of the original Thin-ResNet~\cite{2019arXiv190210107X} is skipped in order to adjust for the specific input shape of the MAGIC images.}. In each of the residual blocks, we deploy a dual squeeze-and-excitation (SE) attention mechanism~\cite{Hu:Attention} to focus on the channel
relationship. We perform either particle classification or regression (energy or arrival direction reconstruction) with a fully-connected head (FCH), a traditional multi-layer perceptron (MLP) neural network. The properties (type, energy, and arrival direction) of the primary particle generating the air shower are reconstructed in the single-task learning mode (see~\cite{Vuillaume:LSTreal} for an IACT-based multi-task learning architecture), where each task is trained with a separate network. We explore stereoscopic information by concatenating the images (integrated pixel charges and signal arrival times) of the two MAGIC telescopes channel-wise before feeding the network as depicted in Fig.~\ref{fig:MAGICDLTRNmodel}.

\subsection{Validation on simulations}

\begin{figure}[ht]
    \centering
    \begin{subfigure}{.32\textwidth}
        \includegraphics[width=\textwidth]{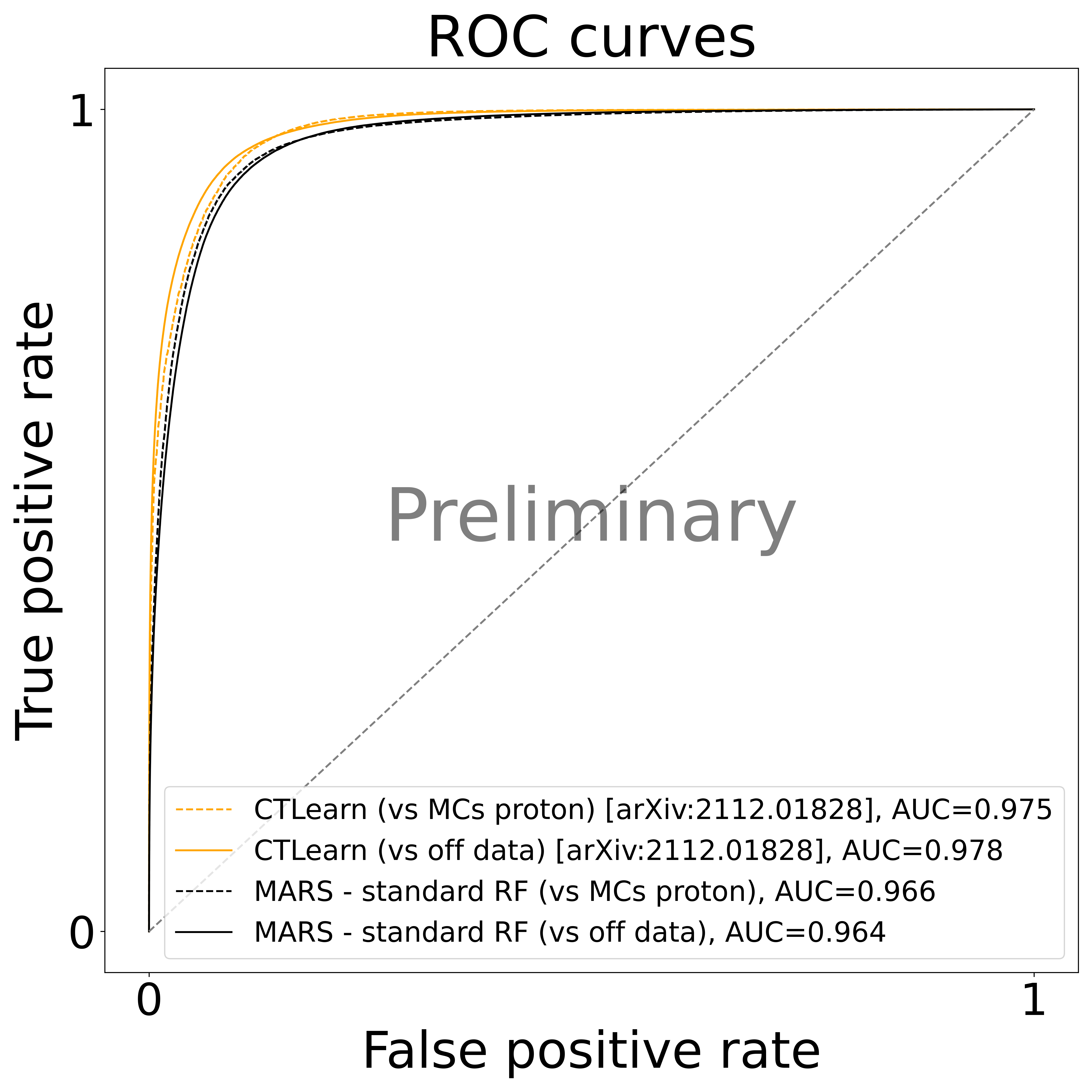}
    \end{subfigure}
    \begin{subfigure}{.32\textwidth}
        \includegraphics[width=\textwidth]{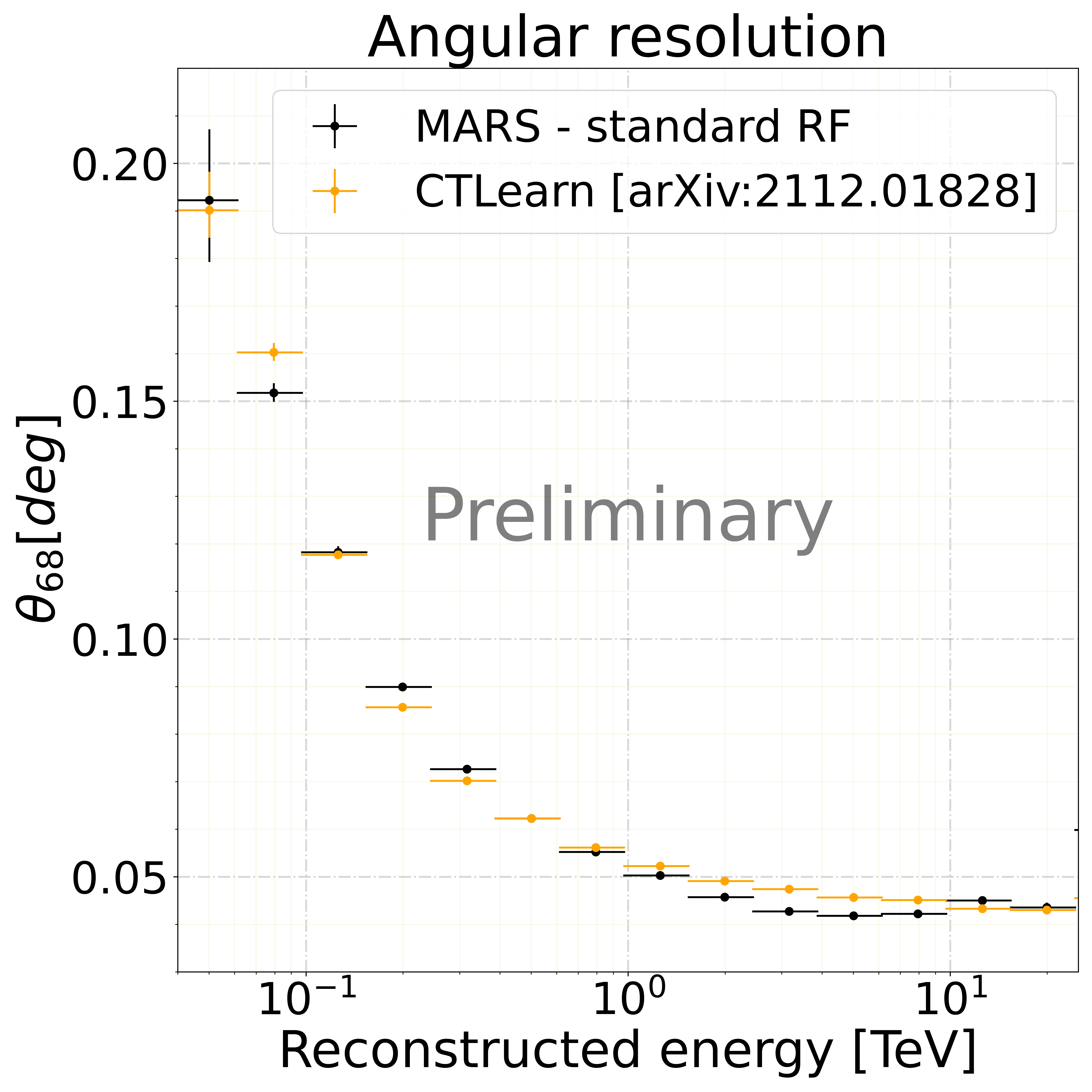}
    \end{subfigure}
    \begin{subfigure}{.32\textwidth}
        \includegraphics[width=\textwidth]{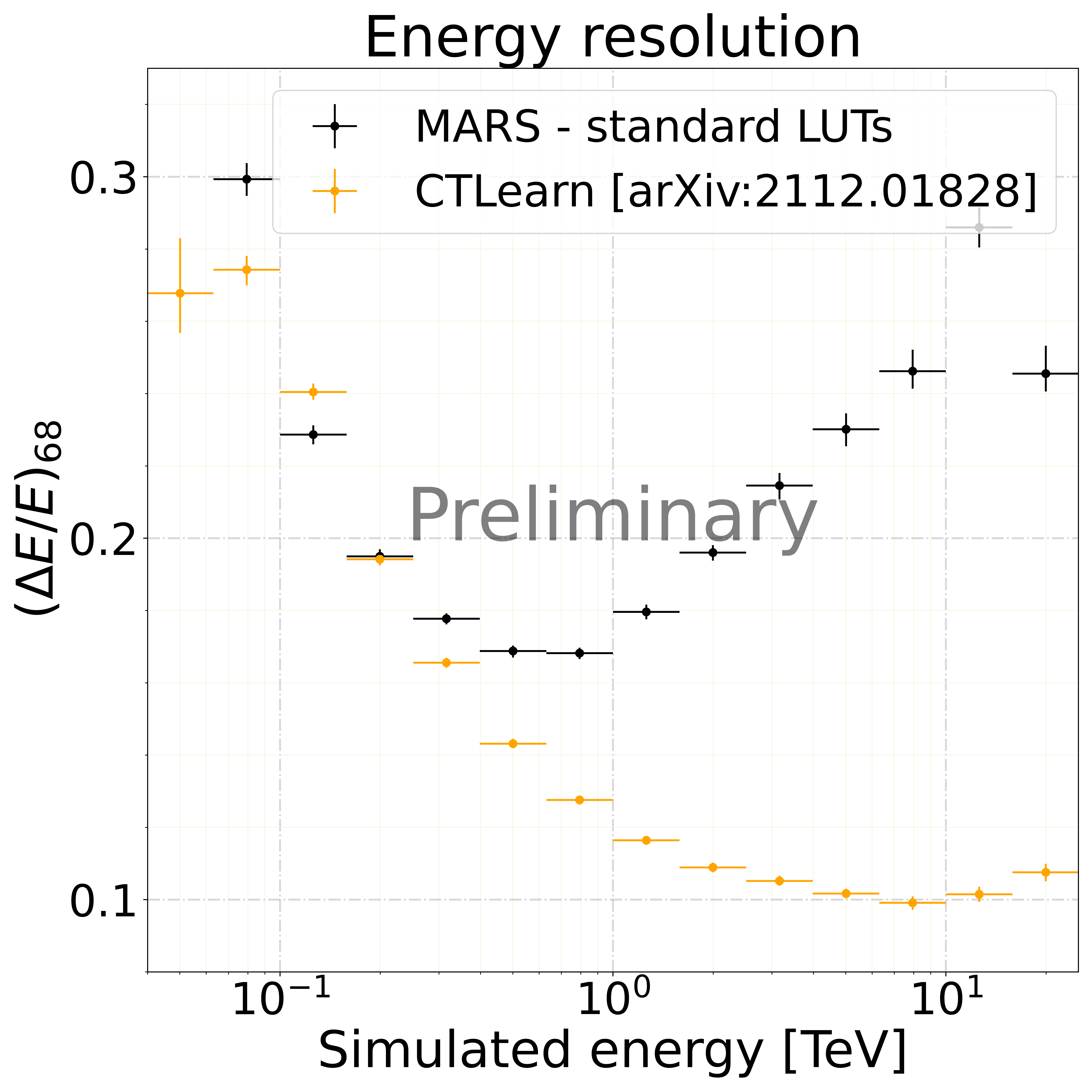}
    \end{subfigure}
\caption{The validation of the performance is taken from~\cite{2021arXiv211201828M}. \emph{Left)} ROC curves with MC proton simulations and observational off data. \emph{Center)} Angular resolution vs. reconstructed energy. \emph{Right)} Energy resolution vs. simulated energy.}
\label{fig:MAGICDLSimulationValidation}
\end{figure}

The evaluation of the performance using common metrics like ROC curves, energy and angular resolution curves with the same quality cuts (see Fig.~\ref{fig:MAGICDLSimulationValidation}) are taken from~\cite{2021arXiv211201828M}. A similar performance is also observed with \texttt{CTLearn v0.6}. Monte Carlo (MC) gamma simulations coming uniformly from a $ 0.4^{\circ} $ offset of the telescope pointing (ringwobble) are used to obtain the reconstruction performance. For the background rejection (see \emph{left panel} of Fig.~\ref{fig:MAGICDLSimulationValidation}), we tested against MC proton simulations and observational off-source data, where we do not expect any gamma-ray signal.

\subsection{Results on observational data}
\begin{figure}[ht]
    \centering
    \includegraphics[width=0.6\textwidth]{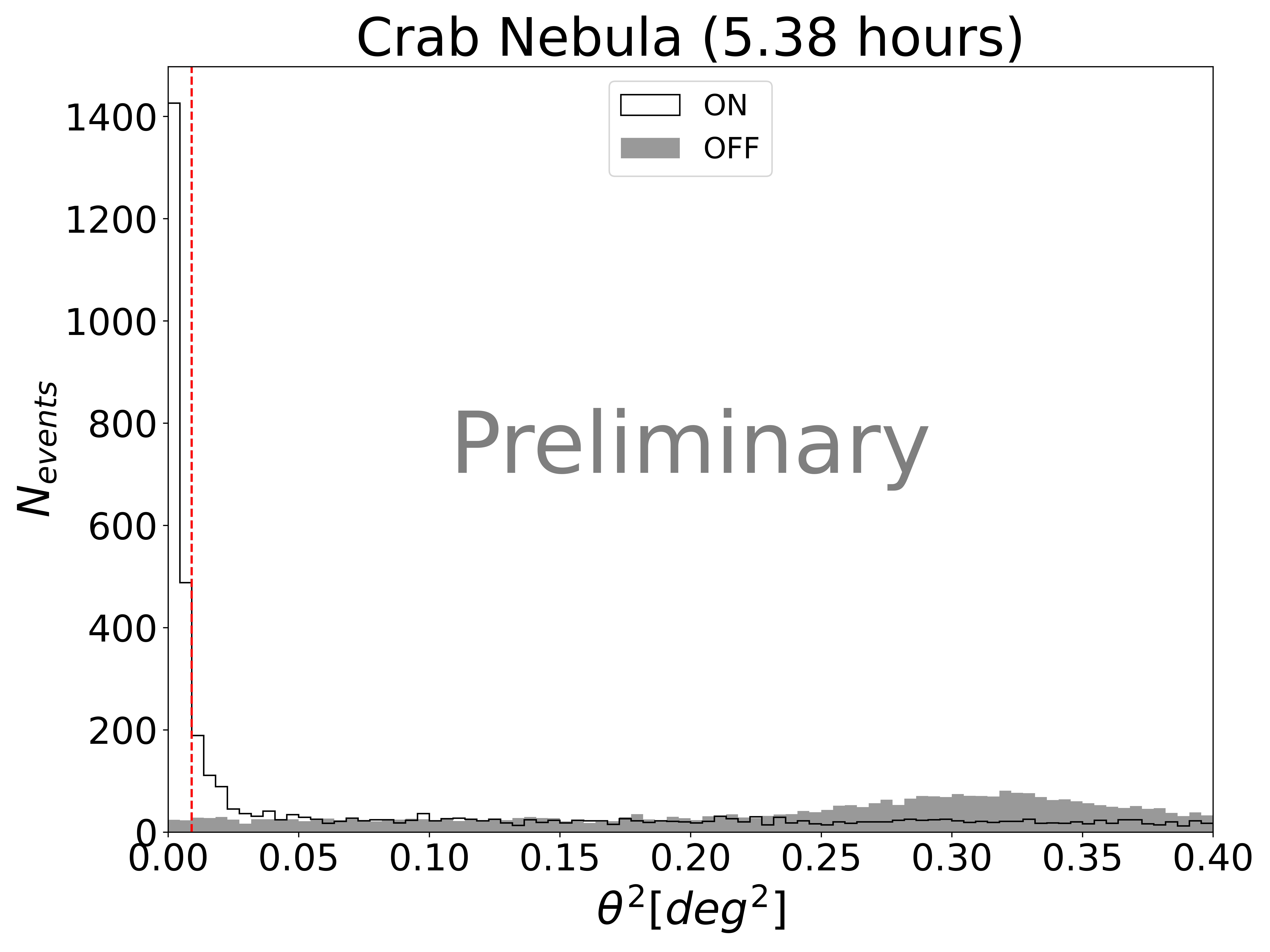}
    \caption{$\theta^{2}$ plot for the \texttt{CTLearn} ME analysis.}
    \label{fig:MAGICDLtheta2CTLearnME}
\end{figure}

We analyzed 5.38 h of observations of the standard gamma-ray candle the Crab Nebula, taken with the MAGIC telescopes on four different nights in 2016 under good weather conditions at low zenith distance (zd < $35^{\circ}$). We used \texttt{MARS} and \texttt{CTLearn} with two settings of analysis cuts (in background suppression and reconstructed energy) focusing on the medium energy (ME; E > $250$ GeV) and low energy (LE; E > $100$ GeV) range. For a fair comparison between the different analysis methods, the background (bkg) rates of the \texttt{CTLearn} analyses are adjusted, through a fine-tuning of the background suppression cut, to match for the corresponding standard \texttt{MARS} analyses (ME or LE). The Crab Nebula is detected using $\theta^{2}$ plots (see Fig.~\ref{fig:MAGICDLtheta2CTLearnME} for the \texttt{CTLearn} ME analysis), where $\theta$ is the angular separation of the source position and the reconstructed arrival direction of the very high-energy photon. The main results of all analyses are summarized in Tab.~\ref{tab:MAGICDLResultsSummary}. The same arrival direction cuts, which defines the fiducial gamma-ray signal region in the $\theta^{2}$ plots, are applied to all different analysis methods. Three off-source positions are considered to evaluate the background distributions. The sensitivity is computed as the strength of the source that gives excess/sqrt(background) = 5 after 50h with the condition of excess/background > 5\% and is given in percentage of the Crab Nebula flux. The significance is calculated following Li\&Ma~\cite{LiMa:1983}.

\begin{table}[ht]
\centering
\resizebox{1\textwidth}{!}{
    \begin{tabular}{|c|c|c|c|c|c|c|c|} 
        \hline
        Analysis & $ N_{on} $ & $ N_{off} $ & $ N_{ex} $ & $ \gamma $ rate [/min]& bkg rate [/min] & Sen. [\% Crab] & Sig. (Li\&Ma) \\
        \hline
        \hline
        \texttt{MARS} – ME & $ 1934 $ & $45.3\pm3.9$ & $1888.7\pm44.1$ & $5.85\pm0.14$ & $0.140\pm0.012$ & $0.58\pm0.03$ & $66.6\sigma$\\
        \hline
        \texttt{CTLearn} – ME & $ 1907 $ & $46.0\pm3.9$ & $1861.0\pm43.8$ & $5.77\pm0.14$ & $0.143\pm0.012$ & $0.60\pm0.03$ & $66.0\sigma$\\
        \hline
        \hline
        \texttt{MARS} – LE & $ 7933 $ & $1827.3\pm24.7$ & $6105.7\pm92.4$ & $18.91\pm0.29$ & $5.661\pm0.076$ & $1.50\pm0.01$ & $83.7\sigma$\\
        \hline
        \texttt{CTLearn} – LE & $ 7889 $ & $1826.3\pm24.7$ & $6062.7\pm92.2$ & $18.78\pm0.29$ & $5.658\pm0.076$ & $1.51\pm0.01$ & $83.2\sigma$\\
        \hline
     \end{tabular}}
    \caption{Summary of all performed analyses (LE/ME and \texttt{MARS}/\texttt{CTLearn}) of the same Crab Nebula sample.}
    \label{tab:MAGICDLResultsSummary}
\end{table}

\section{Conclusions and outlook}
This contribution shows that CNN-based full-event reconstruction works for MC simulations and observational data of the MAGIC telescopes. The performance obtained with \texttt{CTLearn v0.6} matches the sensitivity of detection of the conventional analysis on real data. The selected TRN model is relatively shallow and further performance enhancements are foreseen by increasing the model depth/complexity. We plan to evaluate the full performance of the MAGIC telescopes with CNN-based analyses under various observation conditions in the future.

\section*{Acknowledgments}
\footnotesize We would like to thank the Instituto de Astrofísica de Canarias for the excellent working conditions at the Observatorio del Roque de los Muchachos in La Palma. The financial support of the German BMBF, MPG and HGF; the Italian INFN and INAF; the Swiss National Fund SNF; the ERDF under the Spanish Ministerio de Ciencia e Innovación (MICINN) (FPA2017-87859-P, FPA2017- 85668-P, FPA2017-82729-C6-5-R, FPA2017-90566-REDC, PID2019-104114RB-C31, PID2019-104114RB-C32, PID2019- 105510GB-C31C42, PID2019-~107847RB-C44, PID2019-107988GB-C22); the Indian Department of Atomic Energy; the Japanese ICRR, the University of Tokyo, JSPS, and MEXT; the Bulgarian Ministry of Education and Science, National RI Roadmap Project DO1-268/16.12.2019 and the Academy of Finland grant nr. 317637 and 320045 are gratefully acknowledged. This work was also supported by the Spanish Centro de Excelencia “Severo Ochoa” SEV-2016- 0588, SEV-2017-0709 and CEX2019-000920-S, and "María de Maeztu” CEX2019-000918-M, the Unidad de Excelencia “María de Maeztu” MDM-2015-0509-18-2 and the "la Caixa" Foundation (fellowship LCF/BQ/PI18/11630012), by the Croatian Science Foundation (HrZZ) Project IP-2016-06-9782 and the University of Rĳeka Project 13.12.1.3.02, by the DFG Collaborative Research Centers SFB823/C4 and SFB876/C3, the Polish National Research Centre grant UMO-2016/22/M/ST9/00382 and by the Brazilian MCTIC, CNPq and FAPERJ.
TM acknowledges support from PID2019-104114RB-C32. JLC and DN acknowledges partial support from The European Science Cluster of Astronomy \& Particle Physics ESFRI Research Infrastructures funded by the European Union’s Horizon 2020 research and innovation program under Grant Agreement no. 824064. We acknowledge the support of NVIDIA Corporation with the donation of a Titan X Pascal GPU used for part of this research.
\\
\\
This paper has gone through internal review by the MAGIC Collaboration.


\footnotesize
\begin{thebibliography}{99}

\bibitem{Brill:CTLearn} \textbf{Brill et al.} \texttt{CTLearn v0.6.0}: Deep learning for imaging atmospheric Cherenkov telescopes event reconstruction,  \Journal{\ZENODO}{[10.5281/zenodo.6842323]}{}{2022}.

\bibitem{2017arXiv170905889N} \textbf{Nieto et al.} Exploring deep learning as an event classification method for the Cherenkov Telescope Array, \Journal{\ICRCD}{301}{809}{2017}.

\bibitem{2019ICRC...36..752N} \textbf{Nieto et al.} CTLearn: Deep Learning for Gamma-ray Astronomy, \Journal{\ICRCA}{358}{752}{2019}.

\bibitem{2022ASPC..532..191N} \textbf{Nieto et al.} Reconstruction of IACT events using deep learning techniques with CTLearn, \Journal{\ADASSA}{532}{191}{2022}.

\bibitem{Kim:DL1DH} \textbf{Kim et al.} \texttt{DL1-Data-Handler v0.10.8}: DL1 HDF5 writer, reader, and processor for IACT data, \Journal{\ZENODO}{[10.5281/zenodo.7053921]}{}{2022}.

\bibitem{Zanin:MARS} \textbf{Zanin et al.} MARS, The MAGIC Analysis and Reconstruction Software, \Journal{\ICRCC}{}{773}{2013}.

\bibitem{Aleksic:MAGIC} \textbf{Aleksi\'{c} et al.} The major upgrade of the MAGIC telescopes, Part II: A performance study using observations of the Crab Nebula, \Journal{\ASPP}{72}{76}{2016}.

\bibitem{2021arXiv211201828M} \textbf{Miener et al.} IACT event analysis with the MAGIC telescopes using deep convolutional neural networks with CTLearn, \Journal{\ADASSB}{[arXiv:2112.01828]}{}{2021}.

\bibitem{Chollet:Keras} \textbf{Chollet et al.} \texttt{Keras}, \textbf{https://keras.io} (2015).

\bibitem{tensorflow_developers_2022_5949125} \textbf{TensorFlow Developers} \texttt{TensorFlow v2.8.0}, \Journal{\ZENODO}{[10.5281/zenodo.5949125]}{}{2022}.

\bibitem{Pivarski:uproot} \textbf{Pivarski et al.} \texttt{scikit-hep/uproot4}: 4.1.4, \Journal{\ZENODO}{[10.5281/zenodo.5567737]}{}{2021}.

\bibitem{2019ICRC...36..753N} \textbf{Nieto et al.} Studying Deep Convolutional Neural Networks With Hexagonal Lattices for Imaging Atmospheric Cherenkov Telescope Event Reconstruction, \Journal{\ICRCA}{358}{753}{2019}.

\bibitem{Grespan:LST} \textbf{Grespan et al.} Deep-learning-driven event reconstruction applied to simulated data from a single Large-Sized Telescope of CTA, \Journal{\ICRCB}{395}{771}{2021}.

\bibitem{Miener:CTA} \textbf{Miener et al.} Reconstruction of stereoscopic CTA events using deep learning with CTLearn, \Journal{\ICRCB}{395}{730}{2021}.

\bibitem{2019arXiv190210107X} \textbf{Xie et al.} Utterance-level Aggregation For Speaker Recognition In The Wild, \Journal{\ICASSP}{}{5791}{2019}.

\bibitem{He:ResNet} \textbf{He et al.} Deep Residual Learning for Image Recognition, \Journal{\CVPR}{}{770}{2016}.

\bibitem{Hu:Attention} \textbf{Hu et al.} Squeeze-and-excitation networks, \Journal{\CVPR}{}{7132}{2018}.

\bibitem{Vuillaume:LSTreal} \textbf{Vuillaume et al.} Analysis of the Cherenkov Telescope Array first Large-Sized Telescope real data using convolutional neural networks, \Journal{\ICRCB}{395}{703}{2021}.

\bibitem{LiMa:1983} \textbf{Li and Ma} Analysis methods for results in gamma-ray astronomy, \Journal{\APJ}{272}{317}{1983}.

\end{thebibliography}
\end{document}